\documentclass[aps,prl,preprint,noeprint,superscriptaddress]{revtex4-2}
\usepackage{graphicx}
\usepackage[colorlinks]{hyperref}
\usepackage{natbib}
\usepackage{bm}
\usepackage{amsmath}
\usepackage{color}
\usepackage{bbold}
\usepackage{lipsum}
\usepackage{soul} 

\newcommand{\g}[1]{\mathbf{#1}}
\newcommand{\rev}[1]{\textcolor{black}{#1}}
\setstcolor{red}

\begin{document}

\title{Reflection Measurement of the Scattering Mean Free Path \\ at the Onset of Multiple Scattering}


\author{Antton Go\"icoechea}
\altaffiliation{Present address: Universit\'e de Rennes, CNRS, IETR (Institut d'\'Electronique et des Technologies du num\'eRique), UMR–6164, Rennes, France}
\affiliation{Institut Langevin, ESPCI Paris, PSL University, CNRS, 75005 Paris, France}

\author{C\'{e}cile Br\"utt}
\affiliation{Safran Tech, Digital Sciences and Technologies Department, 78114 Magny-Les-Hameaux, France}

\author{Arthur Le Ber}
\affiliation{Institut Langevin, ESPCI Paris, PSL University, CNRS, 75005 Paris, France}

\author{Flavien Bureau}
\affiliation{Institut Langevin, ESPCI Paris, PSL University, CNRS, 75005 Paris, France}

\author{William Lambert}
\affiliation{SuperSonic Imagine, 13290 Aix-en-Provence, France}

\author{Claire Prada}
\affiliation{Institut Langevin, ESPCI Paris, PSL University, CNRS, 75005 Paris, France}

\author{Arnaud Derode}
\affiliation{Institut Langevin, ESPCI Paris, PSL University, CNRS, 75005 Paris, France}

\author{Alexandre Aubry}
\email[]{alexandre.aubry@espci.fr}
\affiliation{Institut Langevin, ESPCI Paris, PSL University, CNRS, 75005 Paris, France}


\date{\today}

\begin{abstract}
Multiple scattering of waves presents challenges for imaging complex media but offers potential for their characterization. Its onset is actually governed by {the scattering mean free path $\ell_s$ that provides crucial information on the medium micro-architecture.} Here, we introduce a reflection matrix method {designed to estimate this parameter from the time decay of the single scattering rate.} Our method is first validated by an ultrasound experiment on a tissue-mimicking phantom \rev{before being applied in-vivo to a human liver. This study opens important perspectives for quantitative imaging of heterogeneous media with waves, whether it be for non-destructive testing, biomedical or geophysical applications.}
\end{abstract}


\maketitle



Multiple scattering (MSc) of waves proves to be a captivating subject manifesting itself across all the spectrum of wave physics~\cite{Abrahams1979,Albada1985,Lee1985,Tourin1997,Scheffold1998,Larose2004,Hu2008,Vellekoop2008,Popoff2010,Jendrzejewski2012,Gerardin2014,Hsu2017,Shi2018,Horodynski2022,Yamilov2023}.
In an inhomogeneous medium, a {common approach is to consider} a scattering sample as one realization of a random process. Within this paradigm, the relevant parameter for characterizing wave propagation within an {heterogeneous} medium is the scattering mean free path, denoted as $\ell_s$. This parameter represents the typical distance between successive scattering events. For a time-of-flight $t$ smaller than the corresponding mean free time, $\tau_s=\ell_s/c$ (with $c$, the wave velocity), wave propagation behaves akin to a homogeneous medium, displaying a ballistic trajectory. However, as the time of flight increases, scattering events progressively randomize the direction of wave propagation. The trajectory of the wave can be described as a random walk, and energy transport finds an apt model in diffusion theory {for $t>>\tau_s$}. The significance of $\ell_s$ extends beyond its role as a fundamental quantity dictating the onset of MSc; it also holds paramount importance for characterization purposes. 
\rev{Indeed, $\ell_s$ directly correlates with short-scale fluctuations of the wave speed, thereby allowing for instance: ultrasound quantification of fat in liver~\cite{Suzuki1992,Sasso2010,Karlas2017}, osteoporosis diagnosis~\cite{Aubry2008,Karbalaeisadegh2022} or assessment of edema in the lung~\cite{mohantyCharacterizationLungParenchyma2017};  optical mapping of oxy- and deoxy-hemoglobin concentration in neuroscience~\cite{Boas2004} or cornea transparency in optical coherence tomography (OCT)~\cite{Bocheux2019}; characterization of the subsoil for seismic hazard assessment~\cite{Sato2012,Mayor2018} or of polycrystalline structures for non destructive testing~\cite{Li2015,Shahjahan2014}. This list is, by no means, exhaustive but illustrates the large variety of fields  in which a measurement of $\ell_s$ is crucial.} 

From a theoretical standpoint, the probability density function $I(\mathbf{r},t)$ for the travel time $t$ in a random medium, with $\mathbf{r}$ the {relative position} between the source and the receiver, can be obtained by solving the radiative transfer equation. From an experimental standpoint, the averaged instantaneous intensity of the wave-field can be measured and serves as an estimate of
$I(\mathbf{r},t)$. In the following, the word ``intensity'' will be used in the sense of a travel-time distribution. Inverting experimental measurements of $I(\mathbf{r},t)$ allows for the derivation of a spatial mapping of $\ell_s$ along with other relevant transport parameters.
This methodology aligns with the concept of optical diffuse tomography~\cite{Bal2008,Arridge2009,Durduran2010}. However, the practical implementation of this approach is encumbered by a substantial computational load, and its spatial resolution is {poor, as it scales with the imaging depth}. An alternative experimental strategy focuses on the investigation of the wave-field itself, particularly the exponential attenuation of its ballistic component in a transmission configuration~\cite{Page1996}. The characteristic length scale associated with this attenuation is the extinction length, denoted as $\ell_\textrm{ext}$, which encapsulates both scattering and absorption losses: $\ell_\textrm{ext}=(\ell_s^{-1}+\ell_a^{-1})^{-1}$, where $\ell_a$ represents the absorption length. Hence, the {transmitted wave-field} can not furnish an independent measurement of scattering and absorption. Additionally, the impracticality of a transmission measurement is often evident, as only one side of the medium is typically accessible for most applications.
\begin{figure*}
\centering
\includegraphics[width=15cm]{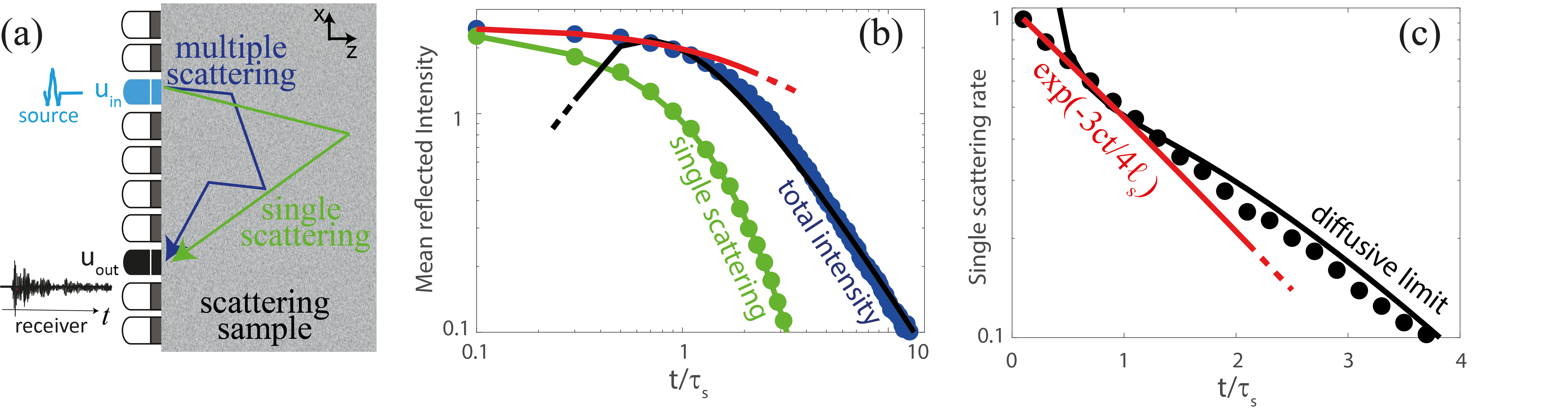}
\caption{\label{theory} (a) Experimental configuration: An array of transducers is placed in front of a 2D semi-infinite random medium [$\ell_s=100$ mm, $\ell_a=100$ m, $c=$1540 m.s$^{-1}$]. (b) Reflected intensity vs. $t/\tau_s$: The single scattering intensity (green dots) obtained via a Monte Carlo simulation of the radiative transfer equation~\cite{supp_mat_ConfocalScat_mfp} is compared with the analytical prediction of Eq.~\eqref{single}; the overall intensity obtained via Monte Carlo (blue dots) is compared with its short-time [Eq.~\eqref{exp}] and long-time [Eq.~\eqref{ssratio_diff}] analytical expressions. 
(c) Single scattering rate $\rho_s(t)$: The numerical result (black dots) is compared with its short-time expression [Eq.~\eqref{ssratio}] and the diffusion model prediction [Eq.~\eqref{ssratio_diff}]. The $x-$ and $y-$axes are in linear and log-scale, respectively.
}
\end{figure*}

A local measurement of $\ell_s$, independent from absorption losses, would be extremely rewarding in reflection, not only for practical reasons, but also for quantitative purposes. In this paper, we demonstrate all these capabilities in the universal framework of matrix imaging. Based on the recording of the reflection matrix associated with a sensor network, it has already found applications in many fields of wave physics ranging \rev{from optical microscopy~\cite{Yoon2020,Badon2020} or ultrasound~\cite{lambertReflectionMatrixApproach2020,velichkoQuantificationEffectMultiple2020} to seismology~\cite{Blondel2018,Touma2022} or microwaves~\cite{Sol2024}. In this paper, the proof-of-concept is performed with ultrasound but the proposed method is very general and can be applied to any kind of wave and sensor network, provided that the spatial sampling of the reflected wave-field satisfies the Nyquist criterion.} Our method is based on a local discrimination between single scattering (SSc) and MSc. Assuming isotropic scattering and solving radiative transfer equation~\cite{paasschensSolutionTimedependentBoltzmann1997}, we first show that the SSc rate $\rho_s(t)$ scales as $\exp [-3ct/(4\ell_s)]$ for $t<\tau_s$\rev{, \textit{i.e} for predominant single scattering ($\rho_s>0.5$)}. This observable can therefore give access to a direct measurement of $\ell_s$. Experimentally, $\rho_s(t)$ is obtained by investigating the reflection matrix of the medium in a focused basis~\cite{lambertReflectionMatrixApproach2020,velichkoQuantificationEffectMultiple2020,Lambert2022} and projecting it onto a characteristic SSc space~\cite{bruttWeightSingleRecurrent2022}.
\rev{The method is first validated on a phantom generating a speckle
mimicking the response of actual tissues to ultrasound.} We provide local measurements of $\ell_s$ and $\ell_a$ in areas showing different scattering properties. We then apply our approach to in-vivo ultrasound data acquired in a liver. In particular, we will show {the} robustness of our method with respect to the variation of reflectivity, in contrast with existing methods relying on attenuation measurements~\cite{Sasso2010,audiereInfluenceHeterogeneitiesUltrasound2013}.

Our approach {applies to} the experimental configuration depicted in Fig.~\ref{theory}(a). An array of $N$ transducers is placed in front of a scattering sample. {These transducers are 10 mm in height, which is much larger than the average wavelength and a vertical cylindrical acoustic lens ensures that the emitted beam remains collimated in the $(x, z)$ plane. When} one element at transverse position $\mathbf{u}_\textrm{in}$ emits an ultrasound pulse, it generates an incident cylindrical wave{,} which is scattered by the heterogeneities of the medium. {Similarly, in reception only waves propagating in the $(x, z)$ plane are recorded by the transducers.} The reflected wave-field $R(\mathbf{u}_\textrm{out},\mathbf{u}_\textrm{in},t)$ is recorded by each transducer identified by its position $\mathbf{u}_\textrm{out}$. By repeating this operation for each element as a source, each reflected wave-field can be stored into a canonical reflection matrix $\mathbf{R}_{\mathbf{uu}}(t)=R(\mathbf{u}_\textrm{out},\mathbf{u}_\textrm{in},t)$. 

Each recorded signal exhibits a complex signature that results {from a random superposition of partial waves, each one being associated with a different scattering path} [Fig.~\ref{theory}(a)]. A  classical approach is to consider this scattering sample as one realization of a random process, and study the overall reflected intensity:
\begin{equation}
I(t) = \mbox{Tr} \left [\mathbf{R}(t)\mathbf{R}^{\dag}(t) \right ] /N,
\end{equation}
where the symbols $\mbox{Tr}$ and $\dag$ stands for matrix trace and transpose conjugate, respectively. The back-scattered intensity $I(t)$ is made of two contributions: (\textit{i}) a SSc component $I_S$ in which the incident wave undergoes only one scattering event before coming back to the sensors {[green arrows in Fig.~\ref{theory}(a)]}; (\textit{ii}) a MSc component $I_M$ that predominates {for $t>>\tau_s$ [blue arrows in Fig.~\ref{theory}(a)]}. To assess the relative weight between each component, radiative transfer equation shall be considered. In a seminal paper, Paasschens~\cite{paasschensSolutionTimedependentBoltzmann1997} solved this equation in a 2D geometry {assuming isotropic scattering}. In particular, he derived an analytical expression for each scattering order of the mean intensity Green's function. Based on this decomposition, {a theoretical expression of $I_S$ can be derived} for a semi-infinite medium~\cite{supp_mat_ConfocalScat_mfp}:
\begin{equation}
\label{single}
I_S(t)=\exp\left (-ct/\ell_\textrm{ext} \right)/({4}\ell_s).
\end{equation}
Not surprisingly, $I_S(t)$ displays an exponential attenuation dictated by $\ell_\textrm{ext}$. As shown by Fig.~\ref{theory}(b), this analytical result is in excellent agreement with the time-of-flight distribution of singly-scattered echoes obtained by means of a Monte Carlo simulation of the radiative transfer equation in a semi-infinite random medium~\cite{supp_mat_ConfocalScat_mfp}. As to the overall intensity, the medium interface can be taken into account in the diffusive approximation~\cite{Patterson1989}. A power law scaling of $I(t)$ as $t^{-3/2}$ is predicted~\cite{supp_mat_ConfocalScat_mfp}:
\begin{equation}
\label{ssratio_diff}
I(t)\underset{t>>\tau_s}{\sim} \frac{\exp(-ct/\ell_a)}{\sqrt{\pi D t}} \frac{z_0}{ct}  
\end{equation}
with $z_0=2\ell_s/3$, the extrapolation length{~\cite{Zhu1991}} and $D$, the diffusion coefficient. In a 2D geometry and for isotropic scattering, $D=c \ell_\textrm{s}/2$. As displayed by Fig.~\ref{theory}(b), this diffusive result predicts well the time dependence of the reflected intensity in the long-time limit but it does not grasp the onset of MSc at short times-of-flight. Interestingly, the radiative transfer solution can provide the asymptotic behavior of $I(t)$ for $t<\tau_s$~\cite{supp_mat_ConfocalScat_mfp}:
\begin{equation}
    \label{exp}
I(t)\underset{t<\tau_s}{\sim}  \exp \left ( {-}ct/4\ell_s\right ) \exp \left (-ct/\ell_a \right)/({4}\ell_s).
\end{equation}
This solution is shown to fit perfectly the time-of-flight distribution of reflected waves obtained by the Monte Carlo simulation at short times-of-flight [Fig.~\ref{theory}(b)]. MSc has thus a strong impact even at small times-of-flight since it modifies the exponential scaling of $I(t)$ with respect to the SSc component [Eq.~\eqref{single}]. This striking property can be highlighted by investigating the SSc rate, $\rho_s(t)=I_S(t)/I(t)$. The ratio of Eqs.~\eqref{single} and \eqref{exp} show{s} that $\rho_s$ displays an exponential attenuation for $t<\tau_s$ that only depends on $\ell_s$:
\begin{equation}
\label{ssratio}
\rho_s(t)\underset{t<\tau_s}{\sim} \exp(-3ct/4 \ell_s).
\end{equation}
Fig.~\ref{theory}(c) illustrates the validity of Eq.~\ref{ssratio} for $t<\tau_s$. {This} fundamental result proves analytically that a discrimination between SSc and MSc can lead to a measurement of $\ell_s$ independent of $\ell_a$~\cite{aubryMultipleScatteringUltrasound2011}. \rev{However, results in Fig.~\ref{theory} are not spatially resolved along the
$x-$axis. A focusing process is required to achieve local measurements.} 
\begin{figure}
\centering
\includegraphics[width=\columnwidth]{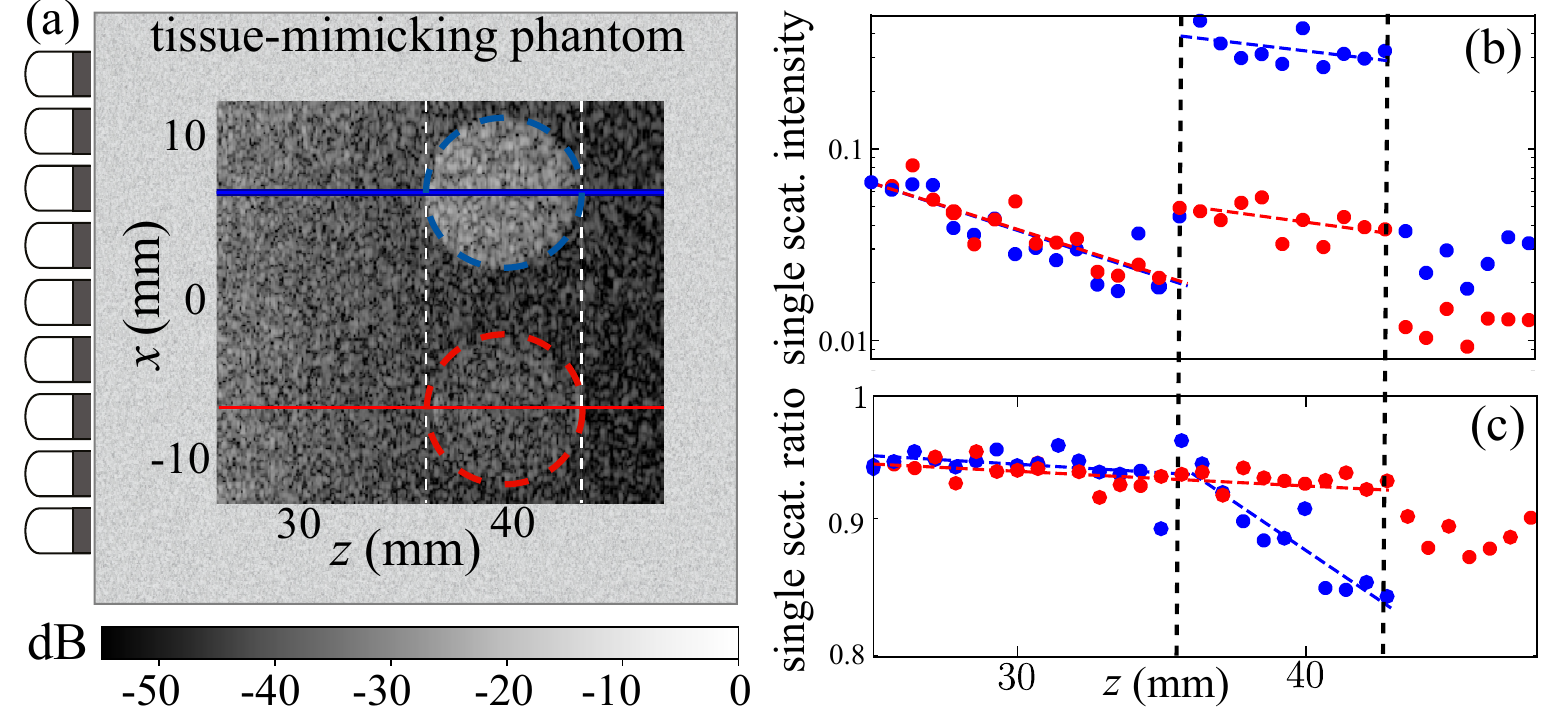}
\caption{\label{phantom exp} (a) Experimental configuration and ultrasound image of the tissue-mimicking phantom. {(b,c) $\hat{I}_S$ (b) and $\hat{\rho}_s$ (c) versus time $t=2z/c$ measured along the blue and red lines displayed in (a). The {$y-$axes} are in log-scale. The experimental measurements are fitted by Eqs.~\eqref{single} and \eqref{ssratio} (dashed lines) in the shallow (25-36 mm) and large (36-43 mm) depth range with values of $\ell_s$ and $\ell_\textrm{ext}$ reported in Tab.~\ref{tab:fit values}.}}
\end{figure}

As a first experimental proof-of-concept, we consider the case of a tissue-mimicking phantom \rev{(CIRS, Model 054GS)} that displays an homogeneous speed of sound $c_0=1540$~mm/$\mu$s and attenuation $\alpha \simeq 0.7$ dB.cm$^{-1}$.MHz$^{-1}$ but also exhibits variations in terms of speckle reflectivity [Fig.~\ref{phantom exp}]. The reflection matrix is acquired using an array of 256 transducers with an inter-element distance of 0.2 mm operating in the 5-10 MHz frequency bandwidth. The first step of the method consists in applying a focused beamforming process to the recorded reflection matrix at input and output~\cite{lambertReflectionMatrixApproach2020,supp_mat_ConfocalScat_mfp}. \rev{A prior optimization of the wave velocity model is performed to limit aberrations~\cite{lambertReflectionMatrixApproach2020}.} The result is a confocal image that is a satisfying estimator of the sample reflectivity under SSc assumption [Fig.~\ref{phantom exp}(a)]. This confocal image shows that the phantom contains two {cylinders} of higher reflectivity than the surrounding speckle [see red and blue dashed lines in Fig.~\ref{phantom exp}(a)]. {However, note that a residual MSc component can still pollute the confocal image.}

\begin{figure}
\centering
\includegraphics[width=8cm]{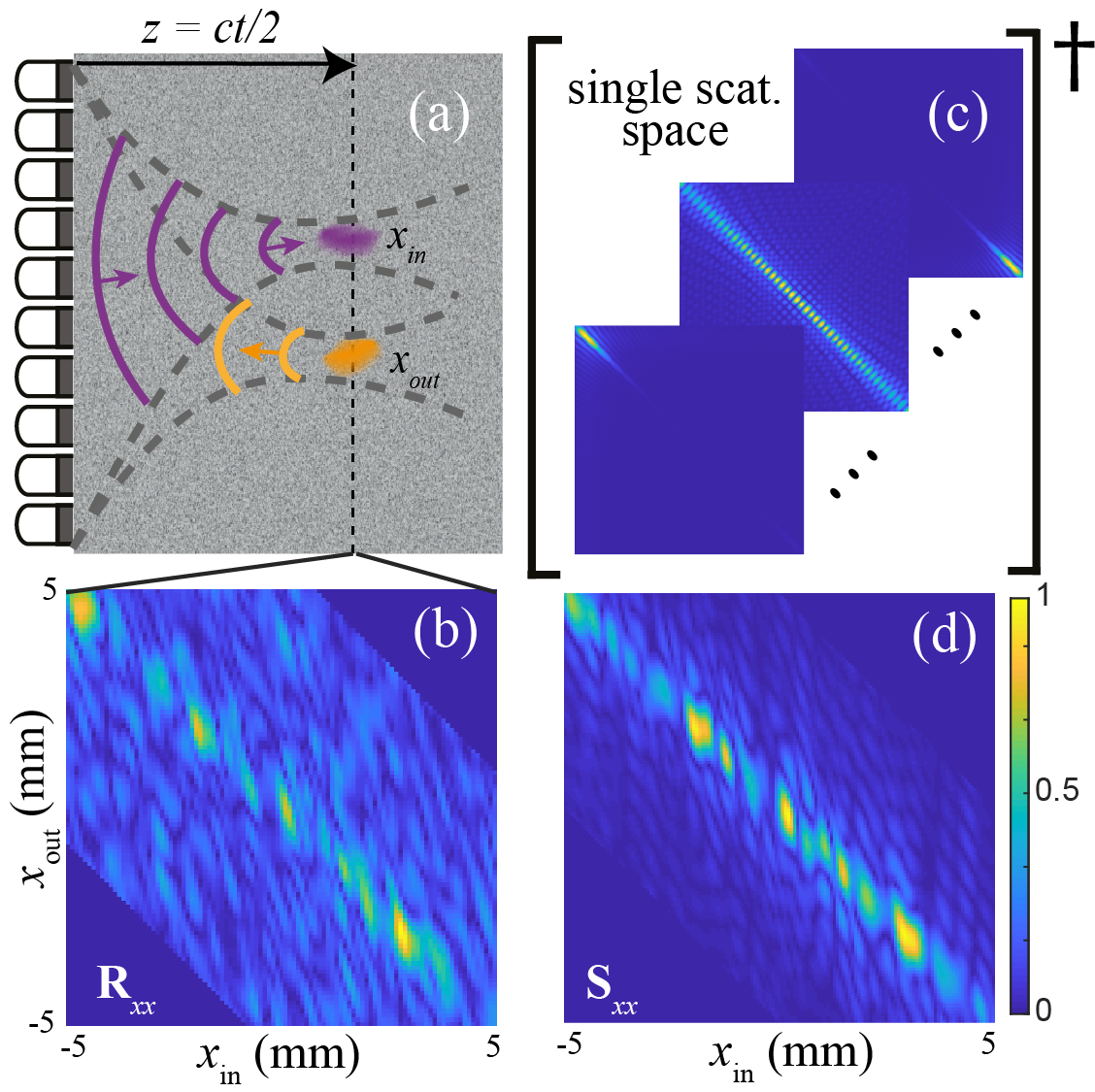}
\caption{\label{schema proc} Principle of the SSc filter. (a) The broadband focused reflection matrix is obtained by focusing in input (purple) and output (orange) at the same depth, yielding $\g{R}_{xx}( z)$. (b) An example of $\g{R}_{xx}(z)$ is displayed at depth $z=45$ mm in the tissue-mimicking phantom. (c) $\g{R}_{xx}(z)$ is projected onto a set of orthonormal matrices forming the so-called {SSc space}~\cite{bruttWeightSingleRecurrent2022}. (d) The result of this projection is a SSc matrix $\g{S}_{xx}(z)$ from which MSc has been discarded.}
\end{figure}
To estimate the weight of MSc, a focused reflection matrix $\mathbf{R}_{xx}(z)$ can be synthesized at each depth $z$~\cite{lambertReflectionMatrixApproach2020}. Its coefficients $R(x_\textrm{out},x_\textrm{in},z)$ are the response of the medium at the ballistic time $t=2z/c$ between virtual sources and detectors located at $(x_\textrm{in},z)$ and $(x_\textrm{{out}},z)$ [see Fig.~\ref{schema proc}(a)].  It displays two contributions: (\textit{i}) A SSc component that emerges along the diagonal or close-diagonal elements; (\textit{ii}) A MSc component that gives rise to a diffuse halo that spreads over off-diagonal elements. To separate both components, an adapted matrix filter has been recently proposed~\cite{bruttWeightSingleRecurrent2022}. It consists in a projection of $\mathbf{R}_{xx}(z)$ onto a set of matrices characteristic of SSc~\cite{supp_mat_ConfocalScat_mfp} [Fig.~\ref{schema proc}(c)]. This operation gets rid {of} the off-diagonal diffuse halo and returns the SSc matrix $\mathbf{S}_{xx}(z)$ [Fig.~\ref{schema proc}(d)]. The norm of $\mathbf{S}_{xx}(z)$ gives access to an estimator of the mean SSc intensity at each depth $z$, or equivalently at each time-of-flight $t=2z/c$:
\begin{equation}
\hat{I}_S(t=2z/c)=\mbox{Tr} \left [\mathbf{S}_{xx}(z)\mathbf{S}^{\dag}_{xx}(z) \right ]/N.
\end{equation}

The single scattering intensity is estimated along the blue and red lines of Fig.~\ref{phantom exp}(a) by placing the probe along each cylinder axis. As expected, $\hat{I}_S$ displays a similar attenuation decay in each region of the phantom whatever the local reflectivity. {At shallow depth}, the measured values for $\ell_\textrm{ext}$ [see Tab.~\ref{tab:fit values}] are in good agreement  with the manufacturer value at the central frequency of 7.5 MHz ($\ell_\textrm{ext} \sim 19$ mm). \rev{In the cylinders, the measured $\ell_\textrm{ext}$ is increased by nearly a factor three but the error bar on this estimation is {important} due to large speckle fluctuations.} 

To go beyond a measurement of $\ell_\textrm{ext}$, one can compute the SSc ratio: $\hat{\rho}_s(t)=\hat{I}_S(t)/I(t)$. Interestingly, the SSc rate seems more robust with respect to the medium reflectivity fluctuations [Fig.~\ref{phantom exp}(c)] \rev{and reduces the uncertainty on the estimation of $\ell_s$ compared with $\ell_\textrm{ext}$}. $\hat{\rho}_s$ exhibits a very different behaviour in the blue and red areas [Fig.~\ref{phantom exp}(c)]. 
\rev{In the blue cylinder, $\ell_s$ is much smaller than its value in the phantom tissue upstream. Not surprisingly, the high contrast exhibited by this cylinder is therefore explained by a stronger scattering. In the red cylinder, $\ell_s$ is the same as in the tissue upstream and almost two orders of magnitude larger than $\ell_\textrm{ext}$. Absorption is therefore the main mechanism accounting for attenuation and the slightly larger contrast exhibited by the red cylinder in Fig.~\ref{phantom exp}(a) is explained by a weaker absorption compared to surrounding speckle. This proof-of-concept experiment thus demonstrates our ability to provide a local discrimination between scattering and absorption losses. } 

\begin{table}
\caption{\label{tab:fit values}Values of $\ell_{\textrm{ext}}$ and $\ell_s$ measured along the blue and red lines of the phantom displayed in Fig.~\ref{phantom exp}.}
\begin{ruledtabular}
\begin{tabular}{c|c|c}
Depth range & $\ell_{\textrm{ext}}$ &  $\ell_s$ \\
\hline
25-36 mm (red) & \rev{17 $\pm$ 2} mm &  \\
36-43 mm (red) & \rev{39 $\pm$ 11} mm  & \rev{1280 $\pm$ 150} mm \\ \cline{1-2}
25-36 mm (blue) & \rev{19 $\pm$ 1 mm} &  \\ \cline{3-3} 
36-43 mm (blue)  & \rev{49 $\pm$ 15} mm & \rev{89 $\pm$ 6} mm   \\
\end{tabular}
\end{ruledtabular}
\end{table}

\begin{figure}
\centering
\includegraphics[width=\columnwidth]{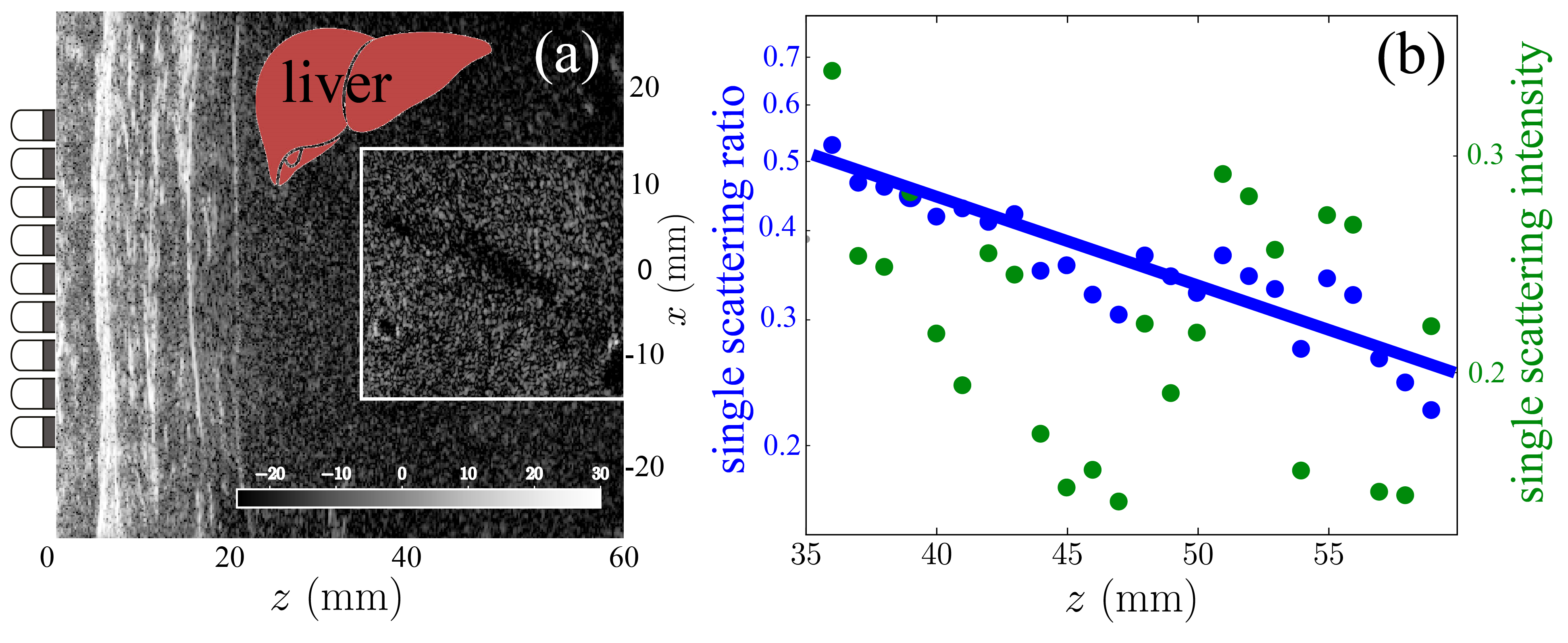}
\caption{\label{liver exp} (a) Experimental configuration and ultrasound image of the liver, including the muscle and fat layers at shallow depths, \rev{with an effective aperture scaling as the imaging depth~\cite{supp_mat_ConfocalScat_mfp}}. The area of interest is surrounded by a white rectangle in which the {gray scale} spans from -25 to 10 dB to improve the image contrast. (b) $\hat{\rho}_s$ (blue dots) and $\hat{I}_S$ (green dots) vs. depth $z=ct/2$. The {$y-$axes} are in log-scale. The fit of $\rho_s(t)$ with Eq.~\eqref{ssratio} (blue line) leads to an estimation of $\ell_s$.}
\end{figure}

Matrix imaging and its SSc filter can therefore provide quantitative markers useful for medical diagnosis. 
As a first step towards such bio-medical applications, we apply our approach to in-vivo ultrasound data acquired on {the} liver of a healthy human subject [Fig.~\ref{liver exp}(a)]. The experimental conditions are the same as for the phantom experiment. The result of our SSc filter is presented in Fig.~\ref{liver exp}(b) for the region surrounded by a white rectangle in Fig.~\ref{liver exp}(a). As previously observed in the phantom, $\hat{\rho}_s$ exhibits much less fluctuations compared to $I_S$. The latter quantity is extremely sensitive to reflectivity variations caused by the presence of structures like veins and cannot provide a reliable measurement of $\ell_\textrm{ext}$. On the contrary, the fitting of ${\hat{\rho}}_s$ with Eq.~\eqref{ssratio} leads to an estimation of $\ell_s \simeq \rev{44 \pm 4}$ mm. This value is remarkably low but is in agreement with the important MSc contribution observed for the same organ in a previous study~\cite{lambertReflectionMatrixApproach2020}. Of course, this demonstration is not limited to liver \rev{nor to ultrasound} and can be extended to any tissue giving rise to speckle. \rev{As a proof for this generality, a recent study has shown how the proposed method can be implemented in OCT for measuring $\ell_s$ in an opaque cornea~\cite{Najar2023}. }

Even though the proposed method is robust with respect to speckle variations, it shows some limitations. First, the SSc filter is only efficient when the MSc background is not too large. The tomography of $\ell_s$ is therefore limited to a penetration depth of a few $\ell_s$. Another issue can arise when wave propagates from a strongly scattering region to a weakly scattering area. MSc induced by the first layer predominates and hides the SSc component of the second one~\cite{supp_mat_ConfocalScat_mfp}. 
A more elaborate inversion procedure of $\rho_s$ is thus needed and will be the object of a future study. {Another relevant question concerns the applicability of this method to a 2D probe, \textit{i.e} a 3D imaging geometry. Interestingly, an exponential decay is still observed at short times-of-flight in 3D: {$\rho_s(t)\sim \exp (-0.55ct/\ell_s)$}~\cite{supp_mat_ConfocalScat_mfp}. Hence the proposed method can be extended to 3D configurations.} {\rev{Another} perspective for this work is its extension to anisotropic scattering. This achievement is particularly crucial for optical microscopy since light scattering is sharply peaked in the forward direction in biological tissues.}

In summary, we have introduced a novel methodology for {local characterization of} scattering media, employing a non-invasive reflection setup. Our approach successfully discriminates between SSc and MSc events, providing an independent measurement of $\ell_s$ and $\ell_a$. This obviates the need for traditional transmission experiments, offering a solution to a longstanding challenge in disordered media research. Beyond its inherent advantages, our method stands as a potent tool for quantitative imaging, applicable to in-vivo or in-situ scenarios. Notably, the robustness of the SSc rate in the face of sample reflectivity fluctuations sets our approach apart from techniques reliant on the time-dependence of backscattered echoes for attenuation measurements. Importantly, while our proof-of-concept experiments were conducted with ultrasound, the universality of matrix imaging renders it applicable to diverse fields of wave physics wherever multi-element technology allows a time-resolved measurement of the reflection matrix~\cite{Balondrade2023,Bureau2023,Giraudat2023}.\\
\vspace{5mm}

\begin{acknowledgments}
All the authors are grateful for the funding provided by the European Research Council (ERC) under the European Union's Horizon 2020 research and innovation program (grant agreement 819261, REMINISCENCE project) and by LABEX WIFI (Laboratory of Excellence within the French Program Investments for the Future, ANR-10-LABX24 and ANR-10-IDEX-0001-02 PSL*). C.B. and W.L. acknowledge financial support from Safran and SuperSonic Imagine, respectively. 
\end{acknowledgments}

\clearpage 

\clearpage

\renewcommand{\thetable}{S\arabic{table}}
\renewcommand{\thefigure}{S\arabic{figure}}
\renewcommand{\theequation}{S\arabic{equation}}
\renewcommand{\thesection}{S\arabic{section}}

\setcounter{equation}{0}
\setcounter{figure}{0}
\setcounter{section}{0}

\begin{center}
\huge{\bf{Supplementary Material}}
\end{center}
\normalsize
\vspace{5 mm}

This Supplementary Material provides further information on: (\textit{i}) the focused reflection matrix; (\textit{ii}) the single scattering filter; (\textit{iii}) the relation between the canonical and focused $\mathbf{R}-$matrices; (\textit{iv}) the theoretical expression of the single scattering intensity; (\textit{v}) the theoretical expression of the overall intensity in the short-time range; (\textit{vi}) the expression of the diffuse intensity in the long-time limit; (\textit{vii}) the numerical validation of the whole method.

\section{Focused reflection matrix}

Matrix imaging basically consists in applying a focused beamforming process to the recorded reflection matrix $\mathbf{R}_{uu}(t)$ at input and output~\cite{lambertReflectionMatrixApproach2020}. To do so, the first step is to perform a temporal Fourier transform of $\mathbf{R}_{uu}(t)$:
\begin{equation}
\label{freq_decomp0}
\overline{\mathbf{R}}_{uu}(\omega)=\int dt \mathbf{R}_{uu}(t) e^{-j\omega t}.
\end{equation}
with $\omega$, the {angular frequency}. The second step is a projection of the reflection matrix in the focused basis. Mathematically, this operation can be performed in the frequency domain by means of the following matrix product:
\begin{equation}
\overline{\mathbf{R}}_{xx} (z,\omega)
=  \mathbf{T}_{ux}^\dag(z,\omega) \times\mathbf{R}_{\mathbf{uu}}(\omega)\times \mathbf{T}_{ux}^{*}(z,\omega).
\label{projRrrcan}
\end{equation}
where the symbols $\dag$ and $*$ stand for transpose conjugation and conjugation operations. $\mathbf{T}_{ux}(z)$ is the transmission matrix that describes wave propagation from the transducer plane ({u}) to the focal plane (x) at each depth $z$. Based on diffraction theory and Rayleigh Sommerfeld integral, its elements $T(u,x,z,\omega)$ correspond to the $z-$derivative of the Green's functions $G(u,x,z,\omega) $ between the transducer and focal planes. $\mathbf{G}_{\mathbf{u} \mathbf{r}}$ depends on the spatial distribution of the wave speed $c(\mathbf{r})$ inside the system. For an homogeneous speed of sound $c_0$ and in a 2D geometry, these Green's functions are given by:
\begin{equation}
G(u,x,z,\omega)=-\frac{j}{4}\mathcal{H}_0\left (k\sqrt{(u-x)^2+z^2} \right )
\end{equation}
where $\mathcal{H}_0$ is the Hankel function of the first kind and $k=\omega/c_0$, the wave number. 

The coefficients $\overline{R}(x_\textrm{out},x_\textrm{in},z,\omega)$  of each focused reflection matrix $\overline{\mathbf{R}}_{xx} (z,\omega)$ are the response of the medium at {angular} frequency $\omega$ between virtual sources and detectors located at $\mathbf{r}_\textrm{in}=(x_\textrm{in},z)$ and $\mathbf{r}_\textrm{out}=(x_\textrm{out},z)$. The last step consists in summing the focused reflection matrix over the frequency bandwidth $[\omega_{-};\omega_{+}]$:
\begin{equation}
\label{freq_decomp}
{\mathbf{R}}_{xx}(z)=\int_{\omega_{-}}^{\omega_{+}} d\omega \overline{\mathbf{R}}_{xx}(z,\omega) e^{j \omega t}.
\end{equation}
{This operation is equivalent to an inverse Fourier transform at lapse time $\tau=0$ in the focused basis; it amounts to a time-gating of back-scattered echoes at the ballistic time $t=2z/c$ in the transducer basis.} It is required to recover the axial resolution of the ultrasound image such that selected singly-scattered waves come from a single plane at depth $z$. 

\section{Single scattering filter}\label{supp_method}
To filter multiple scattering from the focused reflection matrix, an accurate single scattering space should be built~\cite{bruttWeightSingleRecurrent2022}. {To that aim, we first compute a set of monochromatic reflection matrices {${\mathbf{P}}_{uu}(x_s,z,\omega)=[{P}(u_{\textrm{in}},u_{\textrm{out}},x_s,z,\omega)]$ defined in the transducer basis} and associated with a single point-like scatterer located at $(x_s,z)$, such that
\begin{equation}
{\mathbf{P}}_{uu}(x_s,z,\omega)=\mathbf{T}_{ux}(z,\omega) \times \bm{\Gamma}_{xx}(x_s) \times \mathbf{T}_{ux}^{T}(z,\omega) .
\end{equation}
with $\bm{\Gamma}(x_s)$, a diagonal matrix whose coefficients $\gamma(x)$ describes the reflectivity of each scatterer, such that $\gamma(x)=\gamma_0 \delta (x-x_s)$, with $\delta $ the Dirac distribution.}

{The next step is the projection of each matrix $\mathbf{P}_{uu}(x_s,z,\omega)$ in the focused basis, as previously done for $\mathbf{R}_{uu}(z,\omega)$ [Eq.~\eqref{projRrrcan}], 
\begin{equation}
{\mathbf{P}}_{xx}(x_s,z,\omega)=\mathbf{T}_{ux}^{\dag}(z,\omega) \times {\mathbf{P}}_{uu}(x_s,z,\omega) \times \mathbf{T}_{ux}^{*}(z,\omega) .
\end{equation}}
In practice, we generate a synthetic matrix ${\mathbf{P}}_{xx}(x_s,z,\omega)$ for a virtual scatterer every half resolution cell. The dimension of this resolution cell is dictated by the characteristic size $\delta x$ of the focal spot. For an homogeneous speed-of-sound, $\delta x=\lambda/(2\sin [\arctan (z/(2Np))])$.

To build a single scattering space, this set of matrices can be orthogonalized by means a Gram-Schmidt {process}~\cite{bruttWeightSingleRecurrent2022}. However, this approach is not optimal and time consuming. Here we propose a more efficient process based on a singular value decomposition. To that aim, the set of reference matrices is first concatenated into a bi-dimensional matrix $\hat{\mathbf{P}}(z,\omega)$, at each depth $z$ and frequency $\omega$ such that:
\begin{equation}
{\hat{P}}(\lbrace x_\textrm{in},x_\textrm{out} \rbrace , x_s, z,\omega )={P}(x_\textrm{in},x_\textrm{out},x_s,z,\omega)
\end{equation}
Then, the single scattering basis is determined at each depth and frequency by performing a singular value decomposition of $\hat{\mathbf{P}}(z,\omega)$:
\begin{equation}
\hat{\mathbf{P}}(z,\omega)=\sum_{k=1}^{N_k(z,\omega)} \sigma_k (z,\omega)\hat{\mathbf{F}}_k (z,\omega)\mathbf{X}_k (z,\omega)
\end{equation}
where $\sigma_k$ are the singular values ranged in decreasing order. $\hat{\mathbf{F}}_k(z,\omega)=[\hat{F}_k(\lbrace x_\textrm{in},x_\textrm{out} \rbrace,z,\omega)]$ and ${\mathbf{X}}_k(z,\omega)=[X_k(x_s,z,\omega)]$ are the set of singular vectors defined in the basis of focused reflection matrices ($ \lbrace{ x_\textrm{in},x_\textrm{out} \rbrace}$) and the scatterer position basis ($x_s$). \rev{The rank $N_k(z,\omega)$ of the single scattering space is deterministically fixed as the number of resolution cells in the field-of-view:  $N_k(z,\omega)=\Delta x /\delta x(z,\omega)$, with $\Delta x $, the lateral extent of the field-of-view.} 

The set of singular vectors $\hat{\mathbf{F}}_k(z,\omega)$ provide an orthonormal basis of focused reflection matrices ${\mathbf{F}}_{k}(z,\omega)=[F_k(x_\textrm{in},x_\textrm{out},z,\omega)]$ [Fig.2(b) in the accompanying paper] such that
\begin{equation}
F_k(x_\textrm{in},x_\textrm{out},z,\omega)=\hat{{F}}_k(\lbrace x_\textrm{in},x_\textrm{out} \rbrace , z,\omega).
\end{equation}
This single scattering basis is used to project each focused reflection matrix $\overline{\mathbf{R}}_{xx}(z,\omega)$ onto a characteristic single scattering space: 
\begin{equation}
\overline{\mathbf{S}}_{xx}(z,\omega)=\sum_{k=1}^{N_k(z,\omega)} \mbox{Tr} \left \lbrace {\mathbf{F}}_k^{\dag}(z,\omega) \times \overline{\mathbf{R}}_{xx}(z,\omega) \right \rbrace {\mathbf{F}}_k (z,\omega)
\end{equation}
The set of resulting monochromatic matrices $\overline{\mathbf{S}}_{xx}(z,\omega)$ are then recombined to yield the broadband single scattering matrix:
\begin{equation}
\mathbf{S}_{xx}(z)=\int_{\omega_{-}}^{\omega_{+}} d\omega \overline{\mathbf{S}}_{xx}(z,\omega)
\end{equation}
Figure 2(d) of the accompanying manuscript show one example of single scattering matrix in the {phantom} experiment. Not surprisingly, {its diagonal coefficients are significantly larger than off-diagonal coefficients: this is characteristic of single scattering. The ratio between the norms of $\mathbf{S}_{xx}(z)$ and $\mathbf{R}_{xx}(z)$ can serve as an estimator for the single scattering ratio, a refined one compared to previous approaches that estimated the single and multiple scattering rates from the off-diagonal and diagonal intensities~\cite{lambertReflectionMatrixApproach2020}.} Moreover, unlike Ref.~\cite{Lambert2022}, the present filter does not make any assumption on the intensity distribution of multiple scattering.

\rev{Despite these improvements, the single scattering estimator $\rho_s$ can still exhibit a bias $\rho_b$. Indeed, even in absence of single scattering, a residual multiple scattering contribution will still emerge along the single scattering subspace. Assuming the multiple scattering contribution corresponds to
a fully random reflection matrix, one can show the  bias $\rho_b$ is of the order of $1/N_k$. In reality, the multiple scattering intensity profile is of finite spatial extent $W$~\cite{Lambert2022}. Moreover, it exhibits an enhancement by a factor two along the diagonal of the focused reflection matrix due to the coherent back-scattering phenomenon~\cite{lambertReflectionMatrixApproach2020}. Taking into account these two features, the bias is then given by
\begin{equation}
\label{rhob}
\rho_b \sim 2/(1+N_W),
\end{equation}
where $N_W=W/\delta x$ is the ratio between the spatial extent $W$ of the diffuse halo and the resolution cell $\delta x$. Our estimator $\hat{\rho}_s$ of the single scattering ratio is therefore valid only if $\rho_b<<1$. In the phantom and liver experiments described in the accompanying paper, the diffuse halo spreads over the whole matrix ($N_W \sim N$) and the bias $\rho_b$ is negligible.} 

\rev{However, in other configurations, the finite size of the diffuse halo and the coherent back-scattering phenomenon can increase the overlap between the single and multiple scattering sub-spaces. The bias $\rho_b$ might not be negligible but an unbiased estimator $\hat{\rho}'_s$ can be defined as follows:
\begin{equation}
\label{unbiased}
\hat{\rho}'_s=\frac{\hat{\rho}_s - \rho_b}{1 - \rho_b}.
\end{equation}
 A critical step to build this new estimator is to evaluate $\rho_b$. A first option is to compute it by generating random matrices characteristic of the experimental configuration and multiple scattering properties. A second option is to estimate it from the values of $\hat{\rho}_s$ at large echo times for which single scattering is \textit{a priori} negligible. The latter method is actually used in Supplementary Section \ref{sec:num} to build an unbiased estimator of the single scattering ratio for numerical experiments in a far-field configuration.}


\section{Relation between the focused R-matrix and the time-gated canonical R-matrix}

\rev{In the accompanying paper, an equality is assumed between the back-scattered intensity in the transducer basis, $I(t)=\mbox{Tr} \left \lbrace \mathbf{R}_{uu}{(t)} \times  \mathbf{R}_{xx}^{{\dag}}(t) \right \rbrace /N$, and the norm of the focused reflection matrix,  $\mbox{Tr} \left \lbrace \mathbf{R}_{xx}{(z)} \times  \mathbf{R}_{xx}^{{\dag}}(z) \right \rbrace /N$ at the ballistic depth $z=c t/2$ [Eq.~(1)].}

\rev{To show under which approximation this equality is valid,} the first step is to decompose $\mathbf{R}_{xx}$ in the frequency domain [Eq.~\eqref{freq_decomp}] and use the expression of $\overline{\mathbf{R}}_{xx}(\omega)$ given in Eq.~\eqref{projRrrcan} to express the broadband matrix $\mathbf{R}_{xx}$:
\begin{equation}
{\mathbf{R}}_{xx} (z)
= \int d\omega \mathbf{T}_{ux}^\dag(z,\omega) \times\mathbf{R}_{\mathbf{uu}}(\omega)\times \mathbf{T}_{ux}^{*}(z,\omega).
\label{projRrrcan2}
\end{equation}
To simplify the latter expression, the paraxial approximation can be made and the transmission matrix can be decomposed as follows:
\begin{equation}
 \mathbf{T}_{ux}(z,\omega) = \mathbf{F}_{ux}(z,\omega) \exp \left [ j \frac{\omega}{c} z \right ]
\end{equation}
with $\mathbf{F}_{ux}(z,\omega)$, the Fresnel operator whose coefficients $F(u,x,z,\omega)$ are given by:
\begin{equation}
F(\mathbf{u},x,z,\omega) { \simeq e^{3i\pi/4} \sqrt{\frac{2\omega}{\pi z c}}}\exp \left [ j \frac{\omega}{2 z c} (u-x)^2 \right ].
\end{equation}
Using this decomposition of $ \mathbf{T}_{ux}(z,\omega)$, Equation \eqref{projRrrcan2} can be rewritten as follows
\begin{equation}
{\mathbf{R}}_{xx} {(z)}
\simeq \int d\omega \mathbf{F}_{ux}^\dag(z,\omega) \times\mathbf{R}_{\mathbf{uu}}(\omega) \exp \left [ -2j \frac{\omega}{c} z \right ]\times \mathbf{F}_{ux}^{*}(z,\omega).
\label{projRrrcan3}
\end{equation}
To go beyond, the Fresnel operator can be considered as constant over the frequency bandwith and replaced by its value at the central frequency $\omega_c$, such that
\begin{equation}
{\mathbf{R}}_{xx} (z)
\simeq \mathbf{F}_{ux}^\dag(z,\omega_c) \times \mathbf{R}_{\mathbf{uu}}(t=2z/c)  \times \mathbf{F}_{ux}^*(z,\omega_c) .
\label{projRrrcan4}
\end{equation}
\rev{The latter approximation is valid as long as the phase variation of the $\mathbf{F}-$matrix coefficients is negligible over the frequency bandwidth, such that
\begin{equation}
\frac{\Delta \omega D^2}{2zc} <<\pi,
\end{equation}
with $D$, the dimension of the transducer array. This latter equation can be thought of a far-field condition:
\begin{equation}
\label{cond_paraxial}
z>>\frac{\Delta f}{f} \frac{D^2}{\lambda}.
\end{equation}}
The broadband focused reflection matrix ${\mathbf{R}}_{xx} (z)$ and the time-gated reflection matrix $\mathbf{R}_{\mathbf{uu}}(t=2z/c)$ are thus related through a simple change of basis. The unitarity of the Fresnel operator $\mathbf{F}_{ux}(z,\omega_c)$ then implies an equality between their norms:
\begin{equation}
\label{freq_trace}
\rev{I(t)=\mbox{Tr} \left \lbrace {\mathbf{R}}_{uu}(t) \times \overline{\mathbf{R}}^{\dag}_{uu}(t)  \right \rbrace /N \simeq \mbox{Tr} \left \lbrace {\mathbf{R}}_{xx} (z=ct/2)\times {\mathbf{R}}^{\dag}_{xx}(z=ct/2)  \right \rbrace /N  }
\end{equation}
{Upon paraxial approximation,} it is therefore equivalent to compute the back-scattered intensity in the transducer or focused basis. The same relation can be demonstrated for the single scattering contribution: 
\begin{equation}
\label{freq_trace2}
\rev{I_S(t)=\mbox{Tr} \left \lbrace {\mathbf{S}}_{uu}(t) \times \overline{\mathbf{S}}^{\dag}_{uu}(t)  \right \rbrace /N   \simeq \mbox{Tr} \left \lbrace {\mathbf{S}}_{xx} (z=ct/2)\times {\mathbf{S}}^{\dag}_{xx}(z=ct/2)  \right \rbrace /N }
\end{equation}
The analytical expression of $I(t)$ and $I_S(t)$ can thus be derived by considering the response of the medium at its surface. Note that the same result holds if the array of transducers is further away from the sample surface. {In that case, one simply has to set the depth origin ($z=0$) at the sample surface, and accordingly the origin of time ($t=0$) at the arrival time of the first echo.}

\rev{The paraxial approximation of the wave-field is therefore required for a strict equality between the back-scattered intensity and the norm of the focused reflection matrix. Nevertheless, the condition Eq.~\ref{cond_paraxial} is restrictive and can be relaxed by
considering that the characteristic scattering time $\tau_s$ is much larger
than the wave period ($k0\ell_s>>1$ in a weak scattering regime).  To do so, one can consider the relation that exists between the matrices ${\mathbf{R}}_{uu}$ and ${\mathbf{R}}_{xx}$ in the time domain. The coefficients of ${\mathbf{R}}_{xx} (z)$ can also be seen as the result of a delay-and-sum beamforming process:~\cite{Lambert2022}:
\begin{equation}
{R}(x_{\textrm{out}},x_{\textrm{in}},z) = \sum_{u_{\textrm{in}}}\sum_{u_{\textrm{out}}} A(u_{\textrm{out}},  x_{\textrm{out}},z) A(u_{\textrm{in}},  x_{\textrm{in}},z) R(u_{\textrm{out}},u_{\textrm{in}}, 2z/c_0 + {\tau(u_{\textrm{in}},x_{\textrm{in}},z) + \tau(u_{\textrm{out}},x_{\textrm{out}},z)}),
	\label{eq:R_rr__time}
\end{equation}
{where} $\tau$ 
{is} the time delay law such that
\begin{equation}
   \label{eq:timeOfFlight}
\tau(u,x,z) = \lbrace \sqrt{(x -u)^2 + z^2}-z\rbrace /{c_0}{.} 
\end{equation}
$A$ {is} an apodization factor that can 
{limit the extent of} the synthetic aperture and that is normalized such that
\begin{equation}
\sum_{u_{\textrm{in}}}\sum_{u_{\textrm{out}}} A(u_{\textrm{out}},  x_{\textrm{out}},z) A(u_{\textrm{in}},  x_{\textrm{in}},z)=1
\end{equation}
For sake of simplicity, let us consider $A$ as a constant factor ($A=1/N$) and the $\mathbf{R}$-matrix coefficients as fully uncorrelated (random medium). The mean intensity of Eq.~\ref{eq:R_rr__time} can be written as follows:
\begin{equation}
\langle |{R}(x_{\textrm{out}},x_{\textrm{in}},z)|^2 \rangle = \langle |R(u_{\textrm{out}},u_{\textrm{in}}, 2z/c_0 + {\tau(u_{\textrm{in}},x_{\textrm{in}},z) + \tau(u_{\textrm{out}},x_{\textrm{out}},z)})|^2 \rangle .
	\label{eq:I_rr__time}
\end{equation}
The latter equation leads to Eq.~\ref{freq_trace} if the sum of time delays, $\tau(u_{\textrm{in}},x_{\textrm{in}},z) + \tau(u_{\textrm{out}},x_{\textrm{out}},z)$, is negligible compared to the characteristic time over which the mean intensity fluctuates, the mean elastic time $\tau_s=\ell_s/c$. Applying paraxial approximation to Eq.~\ref{eq:timeOfFlight} leads to the following condition for the validity of Eq.~\ref{freq_trace}:
\begin{equation}
z>>\frac{D^2}{\ell_s}
\end{equation}
The latter condition is much less restrictive than Eq.~\ref{cond_paraxial}. Moreover, at shallow depth, it can be forced by applying apodization factors that can reduce the numerical aperture, $D/(2z)$, of the imaging system.}

\section{Single scattering intensity}
\label{ssintensity0}

We first derive an expression for the single-scattering contribution based on radiative transfer theory under isotropic scattering assumption. {In that case, there is no difference between the scattering and the transport mean-free path.} Radiative transfer theory describes the spatial and temporal dependence of the radiance (or specific intensity) $P(\mathbf{r},t,\mathbf{u})$ in a random medium. Radiance is defined as energy flow, propagating in the direction $\mathbf{u}$, per unit normal area per unit solid angle $d \Omega$ per unit time. It follows a transport (Boltzmann) equation :
\begin{equation}
\label{radiative}
\frac{1}{c}\frac{\partial}{ \partial t}P(\mathbf{r},t,\mathbf{u})+\mathbf{u}. \nabla P(\mathbf{r},t,\mathbf{u}) + {\ell_{\textrm{ext}}^{-1}}P(\mathbf{r},t,\mathbf{u}) = {\ell_s^{-1}}P(\mathbf{r},t)+ c^{-1}S(\mathbf{r},t,\mathbf{u}),
\end{equation}
with $S$ the source term {and $P(\mathbf{r},t)$ the intensity, defined as the angular average of the radiance}
\begin{equation}
P(\mathbf{r},t)=\frac{1}{2 \pi}\int d \Omega P(\mathbf{r},t,\mathbf{u})
\end{equation}

Classically, the transport equation can be derived from the Bethe-Salpether equation, neglecting all interference (coherent) effects. Here we adapt the theoretical developments of Paasschens \cite{paasschensSolutionTimedependentBoltzmann1997}, which were derived for an infinite random medium, to our experimental configuration. The problem was solved in two dimensions. In an elegant way, Passchens derived an expression for each scattering order of the intensity Green's function $P(\mathbf{r},t)$. In particular, the single scattering component $P_S$ is derived as follows:
\begin{equation}
\label{single_supp}
P_S(r,t)=\frac{\exp \left ( -ct /\ell_{\textrm{ext}}\right)}{2\pi \ell_s ct} \left ( 1- \frac{r^2}{c^2t^2}\right )^{-1/2} \Theta (ct-r)
\end{equation}
with $\Theta$, the Heav{i}side function: $\Theta (x)=1$ for $x>0$ and zero elsewhere. 

In our case, the medium is semi-infinite. The question is {now how} to modify the result of Paasschens to account for the medium interface. {If we assume that the source term is isotropic, and that the average reflection coefficient for the interface at $z=0$ is 0, we just need to divide by two the result of {Paasschens} for the single scattering component.} The single scattering {intensity} $I_S$ can then be obtained by integrating the mean radiance over the sample surface:
\begin{equation}
I_S(t)=\frac{1}{2}\int_{-\infty}^{+\infty} dx P_S(x,z=0,t)
\end{equation}
Injecting Eq.~\eqref{single} into the last equation leads to the following expression for $I_S(t)$:
\begin{equation}
\label{ssintensity}
I_S(t)=\frac{ \exp (-ct/\ell_\textrm{ext})}{4 \ell_s}
\end{equation}

\begin{figure*}[htbp]
\centering
\includegraphics[width=12 cm]{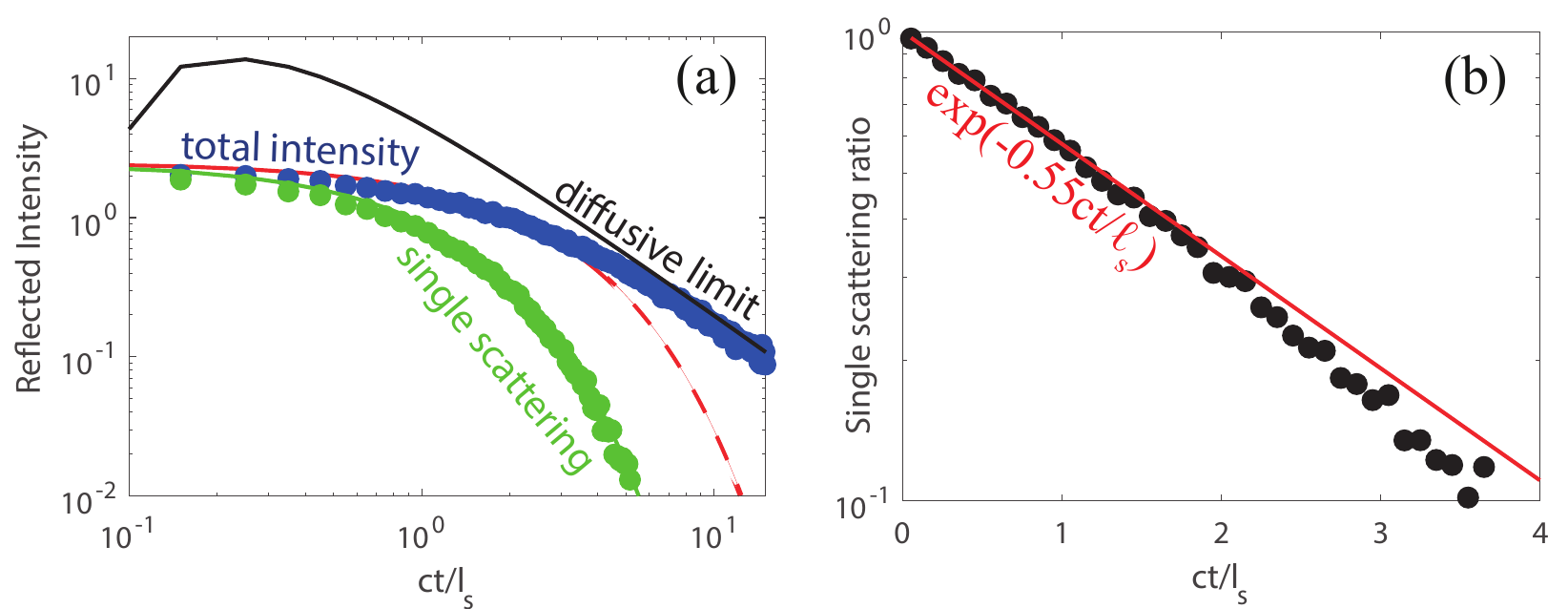}
\caption{\label{fig2} Monte Carlo simulation of energy transport in a 3D semi-infinite random medium [$\ell_s=0.1$ m, $\ell_a=100$ m]. \rev{(a)} Reflected intensity vs. $t/\tau_s$: The single scattering intensity (green dots) is compared with its theoretical expression of Eq.~\eqref{single}; the overall intensity (blue dots) is compared with the heuristic expression of $I(t)=I_S(t)\exp(-0.55 ct/\ell_s)$ (red line) and the 3D diffusive result~\cite{Patterson1989} (black line). \rev{(b)} Single scattering \rev{ratio} $\rho_s(t)$: The numerical result (black dots) is compared with the heuristic exponential decay $\exp(-0.55 ct/\ell_s)$ (red line).}
\end{figure*}

This analytical expression perfectly agrees with the result of the Monte Carlo simulation displayed in Fig.~1(b). {This simulation consists in a 2D random walk with an exponential step size distribution of characteristic length $\ell_s$. $10^6$ particles are thrown in the medium from the origin ($z=0$) with a uniform angular distribution \rev{in the half space ($z>0$)}. The resulting time-dependent backscattered intensity is estimated by building the histogram of times-of-flight at which each reflected particle crosses the interface at $z=0$. This Monte Carlo simulation allows the independent investigation of each scattering order of the backscattered intensity since we know exactly the number of scattering events undergone by each particle before reaching the medium interface.  The single scattering intensity displayed in Fig.~1 has thus been obtained by considering the time-of-flight distribution for particles that have been scattered once and only once before reaching the medium interface. Figure~\ref{fig2}(a) also shows the time-of-flight distribution for the single scattering component but, this time, for a 3D random walk. Interestingly, the expression of $I_S$ given in Eq.~\eqref{ssintensity} also seems to fit pretty well the Monte Carlo result in a 3D geometry [see Fig.~\ref{fig2}(a)]. }

\section{Scattered intensity in the short-time limit}

Paasschens also derived the following analytical expression for the overall radiance $P(\mathbf{r},t)$ {in a 2D geometry}:
\begin{equation}
\label{overall}
P(\mathbf{r},t)=\frac{\exp \left ( -ct /\ell_{a}\right)}{2\pi \ell_s ct} \left ( 1- \frac{r^2}{c^2t^2}\right )^{-1/2} \exp \left [ \left (\sqrt{c^2 t^2-r^2}-ct \right ) /\ell_{s}\right ]  \Theta (ct-r).
\end{equation}
The square root in the exponential can be developed as follows,
\begin{equation}
 \exp \left [ \left (\sqrt{c^2 t^2-r^2}-ct \right )  /\ell_{s} \right ]\sim \exp \left [ -r^2 /(2\ell_{s} ct)\right ]
 \end{equation}
provided that
\begin{equation}
\exp \left [ -r^4 /(8\ell_{s} \rev{c^3}t^3)\right ]  \sim 1.
\end{equation}
Given the fact that $r<ct$, the latter condition is {valid} for $ct<\ell_s$. Equation \eqref{overall} can thus be simplified into the following expression for $t<\tau_s$:
\begin{equation}
\label{overall2}
P(\mathbf{r},t) \underset{t<\tau_s}{\sim} \frac{\exp \left ( -ct /\ell_{a}\right)}{2\pi \ell_s ct} \left ( 1- \frac{r^2}{c^2t^2}\right )^{-1/2} \exp \left [ -r^2 /(2\ell_{s} ct)\right ]  \Theta (ct-r).
\end{equation}

Taking into account the medium interface in presence of multiple scattering is more tricky than for single scattering. A first option is to reproduce what we did for single scattering, i.e dividing by two the mean radiance [Eq.~\eqref{overall}] obtained for an infinite medium. The overall intensity is then given by:
\begin{equation}
I(t)=\frac{1}{2}\int_{-\infty}^{+\infty} dx P(x,z=0,t)
\end{equation}
Injecting Eq.~\eqref{overall2} into the last equation leads to the following expression for $I(t)$:
\begin{equation}
I(t) \underset{t<\tau_s}{\sim}\frac{ \exp (-ct/\ell_\textrm{a})\exp \left [-ct/(4\ell_\textrm{s})\right ]}{4 \ell_s} I_0 \left ( \frac{ct}{4\ell_s} \right)
\end{equation}
where $I_0$ is the modified Bessel function of the first kind. The latter function can be considered as equal to 1 for $t<\tau_s$. The following asymptotic expression is thus obtained for $I(t)$ at short times-of-flight:
\begin{equation}
    \label{exp_supp}
I(t)\underset{t<\tau_s}{\sim}  \exp \left ( - ct/4\ell_s\right ) \exp \left (-ct/\ell_a \right)/(4\ell_s).
\end{equation}
This analytical expression is confronted to the result of the Monte Carlo simulation in Fig.~1(b). A perfect agreement is obtained for $t<\tau_s$ but, as expected, this analytical expression cannot grasp the evolution of the backscattered intensity for $t>\tau_s$. 

Using Eqs.~\eqref{single} and \eqref{exp}, an expression can be derived for the single scattering rate, $\rho_s(t)=I_s(t)/I(t)$, in the short-time limit:
\begin{equation}
\label{ssratio_supp}
\rho_s(t)\underset{t<\tau_s}{\sim} \exp \left (-3 ct/4\ell_s\right ) .
\end{equation}
This is the fundamental result on which the accompanying paper is based. This exponential decay of the single scattering rate is confirmed by the result of the Monte Carlo simulation displayed in Fig.~1(c). Again, an excellent agreement is found between our analytical prediction and the solution of the radiative transfer equation for $t<\tau_s$.  Note that the decay of $\rho_s(t)$ is not strictly identical in a 3D geometry but an exponential decay of $\exp(-0.55 ct /\ell_s)$ fits pretty well the result of our Monte Carlo simulation for $t<\tau_s$ [Fig.~\ref{fig2}(b)].

\section{Diffuse intensity in the long-time limit}

{In a scattering medium ($\ell_s<<\ell_a$), the specific intensity $P(\mathbf{r},t,\mathbf{u})$ can be decomposed as the sum of an isotropic and an anisotropic term~\cite{wang2009}:
\begin{equation}
P(\mathbf{r},t,\mathbf{u})=\frac{\phi(\mathbf{r},t)}{2\pi}+\frac{1}{\pi} \mathbf{J}(\mathbf{r},t).\mathbf{u}
\end{equation}
with $\phi(\mathbf{r},t)$, the fluence rate (or intensity), and $\mathbf{J}(\mathbf{r},t)$, the energy flux. In the long time limit ($t>>\tau_s$), the radiative transfer equation then leads to a much simpler equation for $\phi(\mathbf{r},t)$:
\begin{equation}
\frac{1}{c}\frac{\partial}{\partial t}\phi(\mathbf{r},t) -D \Delta \phi(\mathbf{r},t) + \frac{1}{\ell_a}\phi(\mathbf{r},t)=S(\mathbf{r},t)
\end{equation}
with $D$, the diffusion coefficient. In a 2D geometry and for isotropic scattering, $D=c \ell_\textrm{s}/2$. }

{For an infinite medium,} the Green's function $F$ of the diffusion equation [$S(\mathbf{r},t)=\delta(\mathbf{r})\delta (t)$] can be expressed as follows:
\begin{equation}
F(\mathbf{r},t)=\frac{1}{4\pi Dt} \exp (-ct/\ell_a) \exp \left ( -r^2/ {4Dt}\right ).
\end{equation}
In a semi-infinite medium, the cancellation of $\phi(\mathbf{r},t)$ at the boundary ($z=0$) can be met by adding a negative or image source of energy to the infinite medium problem. The fluence rate per incident {wave packet} can then be written as the sum of contributions from two sources at $z=z_0$ and $z=–z_0$, with $z_0=2\ell_s/3$, the extrapolation length{~\cite{Zhu1991}}. The corresponding Green's function $F(\mathbf{r},t)$ is thus given by:
\begin{equation}
F(\mathbf{r},t)=\frac{1}{4\pi Dt} \exp (-ct/\ell_a) \left [  \exp \left (-\frac{x^2+(z-z_0)^2}{4Dt}\right ) + \exp \left (-\frac{x^2+(z+z_0)^2}{4Dt}\right ) \right ].
\end{equation}
The associated energy flux $\mathbf{J}(\mathbf{r},t)$ at the medium surface can be deduced from Fick's law:
\begin{equation}
\mathbf{J}(\mathbf{r},t)=-(D/c)\nabla F(\mathbf{r},t) |_{z=0}.
\end{equation}
The diffuse intensity, $P_\textrm{d}(\mathbf{r},t)=|\mathbf{J}(\mathbf{r},t)|$, can then be deduced:
\begin{equation}
P_{d}(\mathbf{r},t)=\frac{\exp(-ct/\ell_a)}{2\pi D  t} \frac{z_0}{ct} \exp \left (-\frac{x^2+z_0^2}{4Dt}\right )
\end{equation}
The total diffuse reflectance, $I_d(t)=\int_{-\infty}^{\infty} dx P_d(x,z,t)$, is finally given by:
\begin{equation}
I_{d}(t)=\frac{\exp(-ct/\ell_a)}{\sqrt{\pi D t}} \frac{z_0}{ct}  \exp \left (-\frac{z_0^2}{4Dt}\right )
\end{equation}
In the long-time limit, $4Dt>>z_0^2$. $I_{d}(t)$ thus simplifies into
\begin{equation}
\label{diff}
I_{d}(t)=\frac{\exp(-ct/\ell_a)}{\sqrt{\pi D t}} \frac{z_0}{ct} .
\end{equation}
{Interestingly, this expression of $I_{d}(t)$ is identical to the one previously derived by Patterson \textit{et al.} in 3D~\cite{Patterson1989}}. This expression can be used to fit the result of the Monte Carlo simulation for $t>\tau_s$ both in 2D [Fig.~1(b)] and 3D [Fig.~\ref{fig2}(a)]. If absorption is negligible, a power law decay as $t^{-3/2}$ is thus predicted for the scattered intensity in the diffusive limit.

Using Eqs.~\eqref{single} and \eqref{diff}, an expression can be derived for the single scattering rate, $\rho_s(t)\sim I_s(t)/I_d(t)$, in the long-time limit:
\begin{equation}
\label{diffA}
\rho_s(t)\underset{t>\tau_s}{\sim} \frac{3}{8 } \sqrt{\frac{\pi}{2 }}  (ct/\ell_s)^{3/2} \exp(-ct/\ell_s).
\end{equation}
This diffusion approximation predicts correctly the evolution of the single scattering rate in the long-time limit (Fig.~1c). As in the short-time limit, the single scattering rate can give access, in principle, to a measurement of $\ell_s$ independent from $\ell_a$. Nevertheless, the scaling with $\ell_s$ is more complicated than a simple exponential decay.

\section{Numerical validation}
\label{sec:num}
\begin{figure}
\centering
\includegraphics[width=.8\columnwidth]{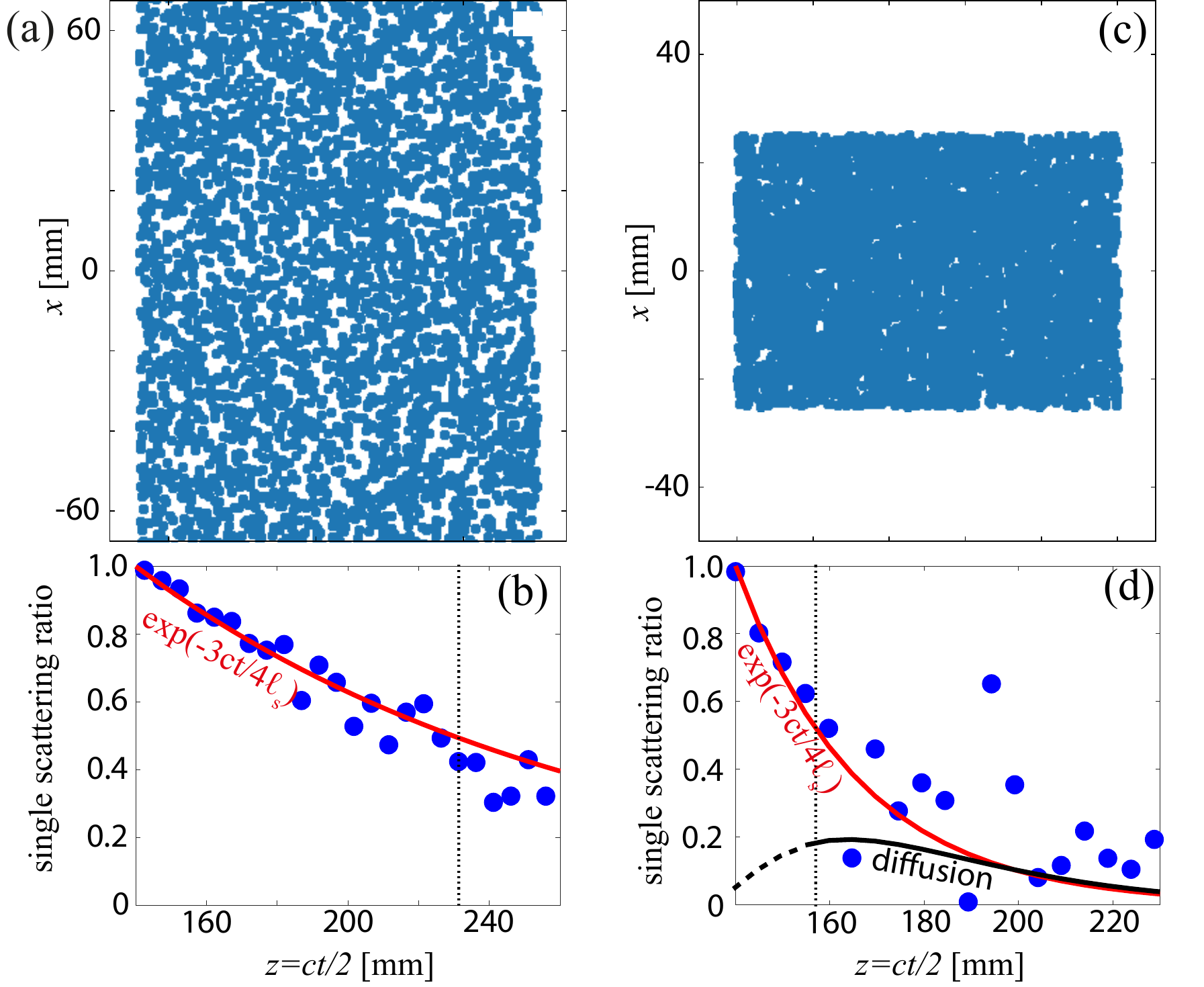}
\caption{\label{coupled dipoles uniform} Single scattering \rev{ratio} for homogeneous disorder. (a) Cylinders uniformly distributed in a 15$\times$13.5 cm$^2$ area with number density $n=0.2$ mm$^{-2}$. $\ell_s$ is estimated from ISA: $\ell_s=194$ mm. (b) Depth-evolution of the unbiased single scattering \rev{ratio $\hat{\rho}'_s$} (blue dots) compared with the theoretical prediction [Eq.~\eqref{ssratio}] using the $\ell_s$ value derived under ISA. (c) Cylinders uniformly distributed in a 15$\times$13.5 cm$^2$ area with number density $n=1$ mm$^{-2}$. The scattering mean free path estimated from ISA is $\ell_s=39$ mm. (d) Depth-evolution of the \rev{unbiased} single scattering \rev{ratio $\hat{\rho}'_s$} (blue dots) compared with \rev{ our theoretical prediction for $t<\tau_s$ [red line, Eq.~\eqref{ssratio}] and the diffusive result [black line, Eq.~\ref{diffA}]} using the $\ell_s$ value derived under ISA. The vertical dashed line accounts for $z=c\tau_s/2$, the boundary between the short and long time regimes.}
\end{figure}

The full method is now validated by means of a simulation of the wave equation. \rev{The scattering medium is a random collection of fluid scatterers ($c_s=2500$ m/s, radius 0.1 mm) embedded in water ($c_0=1480$ m/s) [Fig.~\ref{coupled dipoles uniform}], yielding an individual scattering cross-section  $\sigma=2.58\times 10^{-2}$ mm~\cite{bruttWeightSingleRecurrent2022}. It is assumed that the density contrast between the cylinders and surrounding medium is negligible, so that heterogeneity comes from the compressibility contrast; only longitudinal (pressure) waves are taken into account. The associated reflection matrix is computed over the 1.3 -- 1.7 MHz-frequency bandwidth  by means of a Born series already described in a previous study~\cite{bruttWeightSingleRecurrent2022}. The simulated ultrasound probe is an array of $N=64$ transducers with an inter-element distance $p=0.5$ mm. The array of transducers is placed at 140 mm from the sample surface.}

 In the first system [Fig.~\ref{coupled dipoles uniform}(a)], the concentration of scatterers is $n=0.2$ mm$^{-2}$. Under the independent scattering approximation (ISA)~\cite{shengintroductionwavescattering2006}, the scattering mean free path is given by $\ell_s \sim (n\sigma)^{-1}$, that is to say $\ell_s\sim 194$ mm in the present case. \rev{Under the independent scattering approximation (ISA) that holds in a dilute regime ($k\ell_s>>1$, with $k$ the wave number)~\cite{shengintroductionwavescattering2006}, the scattering mean free path $\ell_s \sim (n\sigma)^{-1}$, that is to say $\ell_s\sim 194$ mm in the present case. ISA is therefore justified ($k \ell_s \sim 1250$) and the value of $\ell_s$ is therefore correct.}

 \rev{The simulated reflection matrix is projected onto the single scattering basis following the method described in Sec.~\ref{supp_method}. Nevertheless, the bias $\rho_b$ (Eq.~\ref{rhob}) is here not negligible because the lateral resolution $\delta x$ is large (far-field configuration). The bias $\rho_b$ is therefore determined from the value of $\rho_s$ at echo times corresponding to ballistic depths beyond the medium thickness. An unbiased estimator is then built using Eq.~\ref{unbiased}.}
 
 The resulting single scattering ratio \rev{$\hat{\rho}'_s(t)$} is displayed in Fig.~\ref{coupled dipoles uniform}(b). A quantitative agreement is found between  \rev{$\hat{\rho}'_s(t)$} and its theoretical scaling in $\exp(-3 ct/4\ell_s)$ in the short time limit. The same observation holds for a more concentrated scattering medium [Fig.~\ref{coupled dipoles uniform}(c)] for which $n=1$ mm$^{-2}$ and $\ell_s\sim 39$ mm. \rev{Again, ISA holds ($k\ell_s \sim 250$)} and our theoretical prediction [Eq.~\eqref{ssratio}] perfectly fits the single scattering rate $\hat{\rho}'_s(t)$ for $t<\tau_s$ [see Fig.~\ref{coupled dipoles uniform}(d)]. This simulation thus validates both the overall method and our theoretical \rev{prediction for $t<\tau_s$. Note, however, that beyond $\tau_s$ (predominant multiple scattering), the estimated single scattering ratio shows strong fluctuations and more spatial averaging would be needed to fit $\rho_s(t)$ with the diffusive prediction of Eq.~\ref{diffA} [black line in Fig.~\ref{coupled dipoles uniform}(d)].}

\begin{figure}
\centering
\includegraphics[width=\columnwidth]{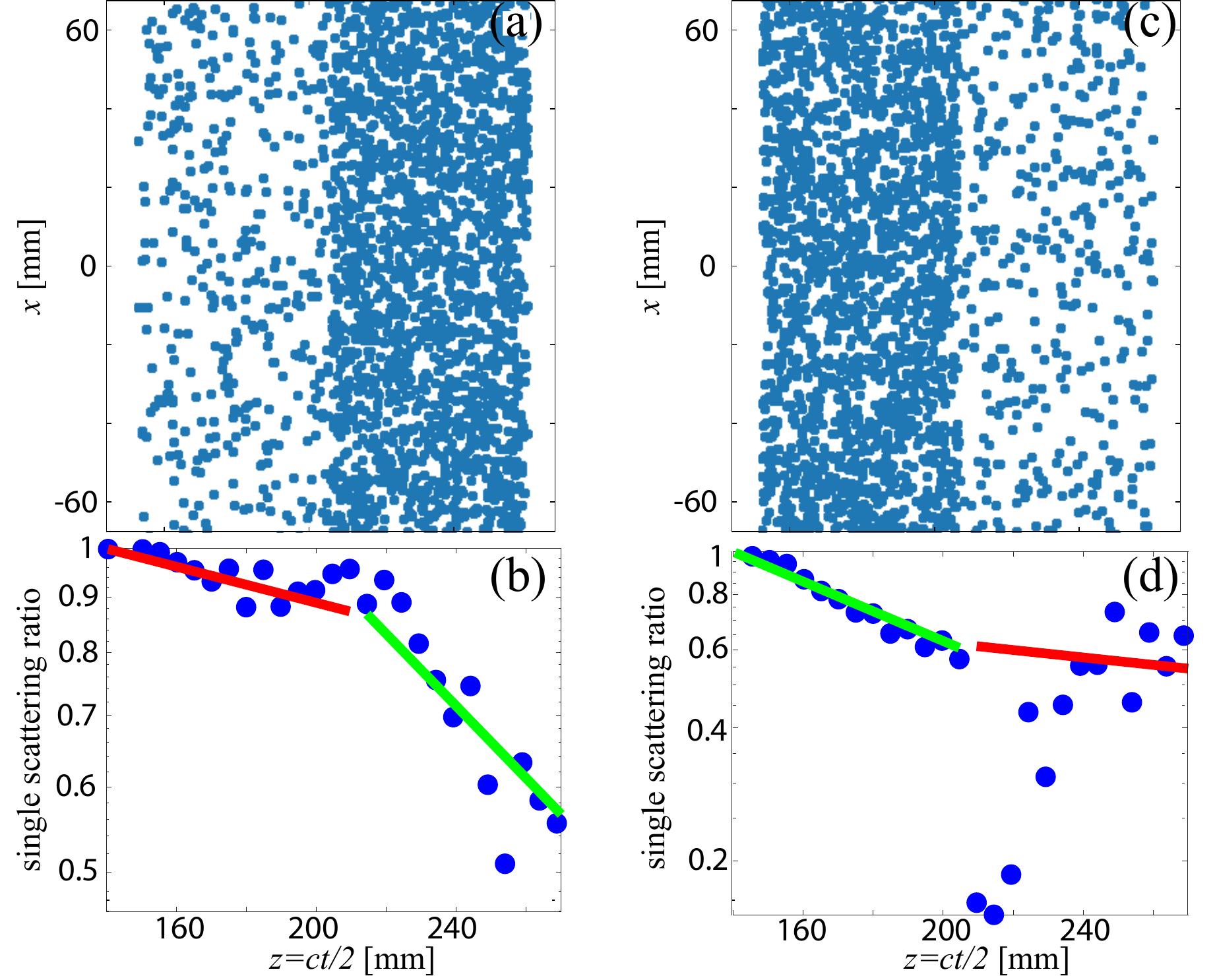}
\caption{\label{coupled dipoles two densities} Single scattering \rev{ratio} for an heterogeneous distribution of disorder. (a) Two-layered system with a dilute-to-dense configuration: $n=0.04$ mm$^{-2}$ ($\ell_s=970$ mm) in the first half of the medium, and dense $n=0.2$ mm$^{-2}$ ($\ell_s=194$ mm) in the second half. (b) Depth evolution of the single scattering \rev{ratio $\hat{\rho}'_s$} (blue dots) compared with the theoretical prediction [Eq.~\eqref{ssratio}] with the $\ell_s$ values derived under the ISA for each scattering layer. (c) Dense-to-dilute system with the same parameters as in (a). (d) Same as in (b) but for the dense-to-dilute system displayed in (c).}
\end{figure}

This numerical simulation can also be used to outline the limits of our approach, in particular in media with {a heterogeneous} distribution of scatterers. To that aim, we now consider a two-layered medium, each layer having a different density of scatterers. Two cases are investigated: dilute-to-dense [Fig.~\ref{coupled dipoles two densities}(a)] and dense-to dilute [Fig.~\ref{coupled dipoles two densities}(c)]. The scatterer concentration in the dilute layer is $n=0.04$ mm$^{-2}$ ($\ell_s \sim 970$ mm), while $n=0.2$ mm$^{-2}$ in the dense layer ($\ell_s \sim 194$ mm). The depth evolution of the \rev{unbiased} single scattering ratio \rev{$\hat{\rho}'_s(t)$} measured in each system is displayed in Figs.~\ref{coupled dipoles two densities}(b) and (d), respectively. As discussed in the main text, propagating from a dilute to a dense region does not affect the single scattering ratio that exhibits a clear change of exponential decay at the interface between the two regions [Fig.~\ref{coupled dipoles two densities}(d)]. On the contrary, in the dense-to-dilute case, \rev{$\hat{\rho}'_s(t)$} completely falls down right after the interface because the multiple scattering coming from the dense region still dominates at times corresponding to ballistic depths located inside the dilute region. Hence, a positive gradient of scattering can be nicely resolved by our method but not a \rev{negative} one. In the latter case, a more subtle inversion of $\rho_s(t)$ is needed to retrieve the depth evolution of the scattering mean free path $\ell_s$.   

%


\end{document}